# A Singularity in the Kirchhoff's Circuit Equations

N R Sree Harsha
Reader in Electromagnetism
*E-mail: nrsreeharshae@gmail.com*

Students often have difficulty in understanding qualitatively the behavior of simple electric circuits. In particular, as different studies have shown, they find multiple batteries connected in multiple loops difficult to analyze. In a recent paper [*Phys. Educ.* **50** 568 (2015)], we showed such an electric circuit, which consists of ideal batteries connected in parallel, that couldn't be solved by the existing circuit analysis methods. In this paper, we shall introduce a new mathematical method of solving simple electric circuits from the solutions of more general circuits and show that the currents, in this particular circuit, take the indeterminate 0/0 form. We shall also present some of the implications of teaching the method. We believe that the description presented in this paper should help the instructors in teaching the behavior of multiple batteries connected in parallel.

## I. INTRODUCTION

Many different studies [1–5] have shown that many of the pre-college and the undergraduate physics students lack a qualitative understanding of the simple electric circuits and were unable answer some of the fundamental questions. In particular, many students found difficult to analyze circuits consisting of multiple batteries connected in single or multiple loops [6, 7]. Despite all these extensive surveys, little attention is drawn to a particular circuit that cannot be solved using any existing circuit analysis methods. In this paper, we shall present the circuit (and hence a class of circuits thereof) and expound the reason for the failure of the circuit analysis methods. The analysis presented in this paper, we believe, should help instructors and students to better understand the behavior of multiple batteries connected in parallel.

Consider the circuit shown in FIG. 1. Let the current from the ideal voltage source $E_1$ be $i_1$ and from the ideal voltage source $E_2$ be $i_2$. In a recent paper [8], we showed that it is not possible to solve for the currents $i_1$ and $i_2$, using the known circuit analysis methods.

In this paper, we shall introduce a new mathematical method of analyzing simple electric circuits in Section II and show that the currents $i_1$ and $i_2$ take the 0/0 form and hence are indeterminate. We shall also present a common misconception in understanding our new method in section III and clarify the misconception in section IV. We shall then present an alternate circuit problem in section V and show that the form of indeterminate is the same.

## II. THE MATHEMATICAL METHOD

We shall introduce the method based on two simple examples. It is important to understand that all the voltage sources in this paper are considered to be ideal and all the resistors are considered to be linear, obeying the Ohm's law. Throughout the paper, we will also be using the terms 'ideal' voltage/current source and 'constant' voltage/current source interchangeably.

For our method, we need to first construct a general circuit as shown in FIG. 2.

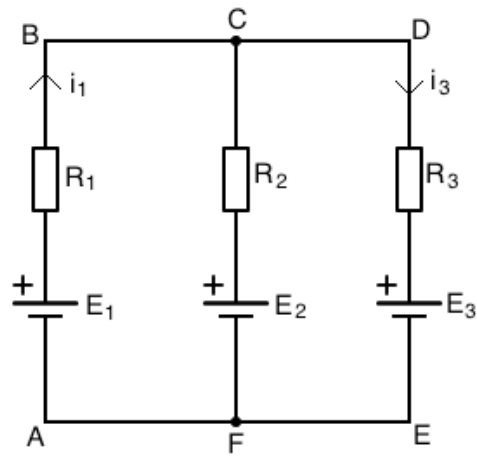

FIG. 2. A general circuit to demonstrate the mathematical method.

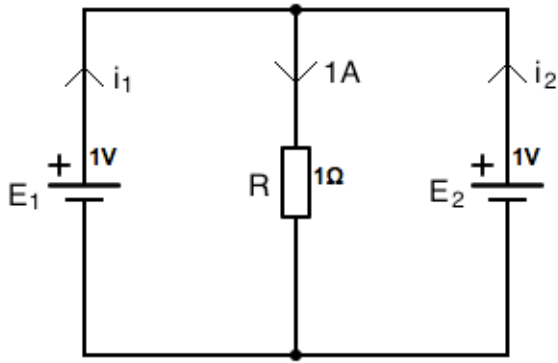

FIG. 1. A simple form of the circuit.

Let the current from the voltage source $E_1$ be $i_1$ and let the current into the voltage source $E_3$ be $i_3$. Applying the Kirchhoff's loop law along the loop ABCF, we get that

$$i_3 = \frac{i_1 R_1 + i_1 R_2 + E_2 - E_1}{R_2} \quad (1)$$

Next, applying the Kirchhoff's loop law along the outer loop ABDE, we see that

$$i_3 = \frac{E_1 - E_3 - i_1 R_1}{R_3} \quad (2)$$

For any combination of resistors and voltage sources of a network, there is only one unique solution of currents that can flow through the network components. Hence, we can safely equate the different values of the current $i_3$ given by the equations (1) and (2). We then get that

$$i_1 = \frac{E_1 R_2 - E_3 R_2 + E_1 R_3 - E_2 R_3}{R_1 R_2 + R_2 R_3 + R_3 R_1} \quad (3)$$

We now have the solution of the current that flows from the voltage source $E_1$ of the circuit shown in FIG. 2. We can now let the values of the voltage sources and resistors of the circuit shown in FIG. 2 vary continuously from zero to infinity. We can thus vary the values of the different network components and arrive at the solution of various other simpler circuits.

### 1. Resistors connected in series

Consider the circuit shown in FIG. 3. The problem is to find out the current $i$ that flows through the two resistors. We can find out the value of the current $i$ by appropriately varying the values of the resistors and voltage sources of the equation (3).

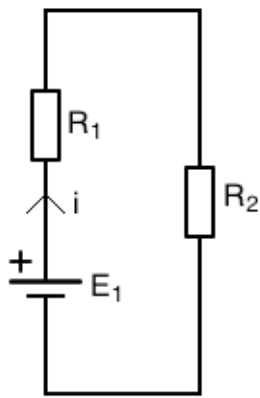

FIG. 3. A simple circuit that can be solved from the solution of the general circuit.

It can be easily understood that as $R_3 \to \infty$ and $E_2 \to 0$, while the other parameters remain constant, the circuit shown in FIG. 2 becomes the circuit shown in FIG. 3. Hence, we can write

$$i = \lim_{\substack{R_3 \to \infty \\ E_2 \to 0}} i_1 = \lim_{\substack{R_3 \to \infty \\ E_2 \to 0}} \frac{E_1 R_2 - E_3 R_2 + E_1 R_3 - E_2 R_3}{R_1 R_2 + R_2 R_3 + R_3 R_1} \quad (4)$$

Next, dividing the numerator and denominator by $R_3$, we get

$$i = \lim_{\substack{R_3 \to \infty \\ E_2 \to 0}} \frac{\frac{E_1 R_2}{R_3} - \frac{E_3 R_2}{R_3} + E_1 - E_2}{\frac{R_1 R_2}{R_3} + R_2 + R_1} \quad (5)$$

Hence, we get

$$i = \frac{E_1}{R_1 + R_2} \quad (6)$$

This is the usual solution obtained by applying the Kirchhoff's laws to the circuit shown in FIG. 3. However, we have obtained equation (6) without applying the Kirchhoff's laws to the circuit shown in FIG. 3.

### 2. Voltage sources in series

We now consider the circuit shown in FIG. 4, where two ideal voltage sources are connected in series.

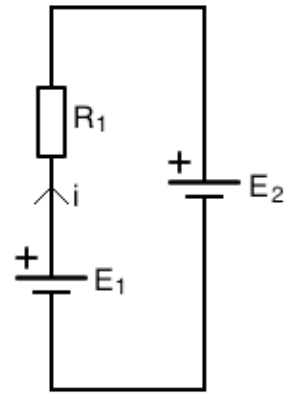

FIG. 4. Another simple circuit that can also be solved from the solution of the general circuit.

We have to find out the current $i$ that flows through the load resistor $R_1$. It is clear that as $R_2 \to 0$ and $R_3 \to \infty$, the circuit shown in FIG. 2 becomes the circuit shown in FIG. 4. Hence, we have

$$i = \lim_{\substack{R_3 \to \infty \\ R_2 \to 0}} \frac{\frac{E_1 R_2}{R_3} - \frac{E_3 R_2}{R_3} + E_1 - E_2}{\frac{R_1 R_2}{R_3} + R_2 + R_1} \quad (7)$$

Which gives us that

$$i = \frac{E_1 - E_2}{R_1} \quad (8)$$

This is what we would get, if we apply the Kirchhoff's laws to the circuit shown in FIG. 4.

Hence, by varying different network parameters of a general circuit, we can obtain the solution to various other circuits. Note, however, that the Kirchhoff's laws are needed to obtain the solution of the general circuit. But, after solving the general circuit once, we can obtain the solution of many simpler circuits just by varying the appropriate network components of the general circuit. This is one of the advantages of our new mathematical method. The second advantage is this: the conventional network analysis methods fail to give us a reason for the failure of analyzing the circuit shown in FIG. 1. But, our new mathematical method, as we shall see in the following sections, tells us the reason why the circuit analysis methods fail to give us an answer.

## III. A MISCONCEPTION

The primary purpose of introducing a new method in the previous section is to teach the students about the behavior of multiple batteries connected in parallel across a load. But, there was a misconception observed among the students, that merits an attention in this paper. We shall first present the misconception in this section.

Consider the general circuit shown in FIG. 5. We have taken two identical voltage sources of *emf* '*E*' with identical internal resistances '*r*'. A load '*R*' is connected to the parallel combination of these sources. Let the current from the sources be $i_1$ and $i_2$ respectively.

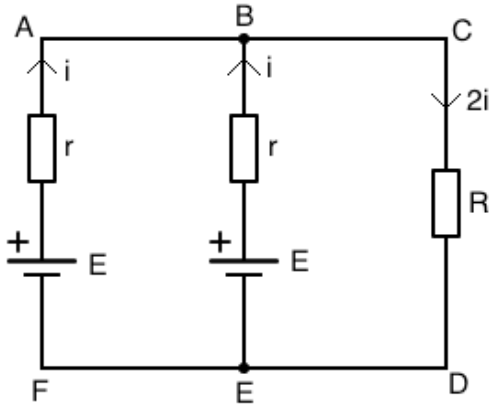

FIG. 5. A circuit to demonstrate the misconception.

Applying the Kirchhoff's loop law in the loop ABEF, we get

$$E - E = r(i_1 - i_2) = 0 \quad (9)$$

This clearly implies that $i_1=i_2$ when $r\neq0$. Thus, the currents are equally distributed when a non-zero finite internal resistance of equal value is present. Let the value of this common current be '*i*' (i.e. $i_1=i_2=i$). Now, applying the Kirchhoff's law in the loop ACDF, we get

$$i = \frac{E}{r + 2R} \quad (10)$$

As we let *r* tend to zero, the circuit shown in FIG. 5 becomes the circuit shown in FIG. 1. The current *i* from each voltage source of the circuit shown in FIG. 1 is then given by

$$i = \lim_{r \to 0} \frac{E}{r + 2R} \quad (11)$$

This implies that $i \to \dfrac{E}{2R}$ as $r \to 0$.

With $E=1$ Volt and $R=1\Omega$ as shown in FIG. 1, it implies that $i=0.5$ A. Thus, the circuit shown in FIG. 1 can be solved by mere application of the Kirchhoff's voltage law.

## IV. AN EXPLANATION

In this section, we shall point out the mistake in the analysis presented in the previous section and show that applying the Kirchhoff's laws cannot solve the circuit shown in FIG. 1. We shall divide this section into two parts.

### 1. Continuity of the functions f(x) and g(x)

In order to present the concepts more clearly, we shall now study the mathematical ideas of the continuity of a function and its relationship with the value of the function.

As an example, consider the two functions $f(x) = \dfrac{x^2}{x}$ and $g(x) = x$.

The domain of the function f(x) is (-∞,0) **U** (0,+∞) while the domain of the function g(x) is (-∞,+∞). Here, '**U**' represents the 'Union' operator. For all our purposes, (..) represents an open interval and [..] represents a closed interval. This implies that the domain of the function g(x) can also be written as (-∞,0] **U** [0,+∞). Thus, the point x=0 is included in the domain of the function g(x), but it is excluded from the domain of the function f(x) as 'f' is undefined at x=0.

Hence, we say $f(x) = \dfrac{x^2}{x} = x$ for $x \neq 0$. This essentially means that if we do not consider the point x=0, we can safely say that $f(x) \equiv g(x)$. It should be clear that the function f(x) is not defined at the point x=0.

Let us now try to find out the limit of the function g(x) as 'x' tends to zero. Clearly,

$$\lim_{x \to 0} g(x) = \lim_{x \to 0} x = 0 \qquad (12)$$

Next, consider the limit of the function f(x) as 'x' tends to zero. We should be more careful at this point. When we define the limit of any function y=h(x) at a point 'x=a', 'a' should either be in the domain of the function 'h(x)' or should coincide with one of the end points of its domain (these end points can either belong to an open interval or a closed interval).

We have

$$\lim_{x \to 0} f(x) = \lim_{x \to 0} \dfrac{x^2}{x} = 0 \qquad (13)$$

Since, the point x=0 coincides with one of the end points of the domain of the function f(x), it is safe to say that the function f(x) tends to zero as x tends to zero. Hence, our statement $f(x) \to 0$ as $x \to 0$ is correct.

Thus, $\lim_{x \to 0} g(x) = \lim_{x \to 0} f(x) = 0$.

Next, consider the value of the function g(x) at the point x=0. This essentially brings us to the idea of the continuity of the function and its relationship with the limit of the function. The question is: when can we say that $\lim_{x \to 0} h(x) = h(0)$ for any general function h(x)? One of the conditions is that the function h(x) should be defined at x=0 [9]. If this condition is not met, we cannot say that $\lim_{x \to 0} h(x) = h(0)$.

With this, it is clear that $\lim_{x \to 0} g(x) = g(0) = 0$.

But, what about the value of the function f(x) at x=0? Certainly, $\lim_{x \to 0} f(x) = 0$. However, the function f(x) is not defined at x=0. This means that $\lim_{x \to 0} f(x) \neq f(0)$. This essentially means that f(0)≠0. Thus, the two functions f(x) and g(x) differ at this crucial point. It is one thing to say that f(x) has a limit at x=0, but quite an other thing to say f(0) is equal to zero, which is incorrect.

2. **The circuit problem**

Let us now consider our circuit under study. The current from each voltage source of the circuit shown in FIG. 5 is given by equation (10). We have obtained equation (10) from the fact that $i_1 = i_2 = i$ when $r \neq 0$. This is obtained from equation (9), which is shown below

$$r(i_1 - i_2) = 0 \qquad (14)$$

We can write the current 'i' defined in equation (10) as i(r) ('i' as a function of 'r'). We then have

$$i(r) = \dfrac{E}{r + 2R} \text{ when } r \neq 0 \qquad (15)$$

But, as seen from equation (14), $i_1$ need not be equal to $i_2$ if r=0. Hence, equation (15) is not justified for the case when r=0. Note, however, that the limit of the function 'i(r)' is E/2R as $r \to 0$. Hence, $i(r) \to E/2R$ as $r \to 0$ but i(0) is not defined. We need the value of i(0) but not the value of the limit of i(r) as r tends to zero.

This is the mistake in the analysis presented in section III. To overcome this problem, let us consider a more general circuit, which does not assume that $i_1 = i_2$ as shown in FIG. 6.

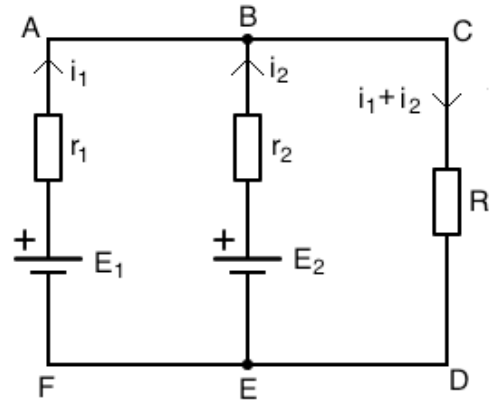

FIG. 6. A correct general circuit.

Applying the Kirchhoff's law in the loop ABEF, we have

$$E_1 - E_2 = i_1 r_1 - i_2 r_2 \qquad (16)$$

Applying the Kirchhoff's law in the loop BCDE, we get

$$E_2 = i_2 r_2 + (i_1 + i_2) R \qquad (17)$$

Eliminating $i_2$ from equations (16) and (17), we get

$$i_1 = \dfrac{E_1 r_2 + E_1 R - E_2 R}{r_1 r_2 + r_1 R + r_2 R} \qquad (18)$$

Since we are interested in a particular case of $E_1=E_2=E$ and $r_1=r_2=r$, we get

$$i_1 = \frac{Er}{r^2 + 2rR} \qquad (19)$$

And, we also have

$$i_2 = \frac{Er}{r^2 + 2rR} \qquad (20)$$

Thus, as seen from equations (19) and (20), $i_1=i_2$ when $r\neq 0$. Let us call this current "$i$".

Since, we are interested in current '$i$' as a function of the internal resistance '$r$' (we shall write it as '$i(r)$'), we can write equation (19) and (20) as

$$i(r) = \frac{Er}{r^2 + 2rR} \qquad (21)$$

Clearly as $r \to 0$, $i(r) \to \frac{E}{2R}$ which confirms with analysis following the equation (15). But, as explained before, the value of the function $i(r)$ is not defined at the point $r=0$. This means that $\lim_{r \to 0} i(r) \neq i(0)$. Hence, $i(0) \neq \frac{E}{2R}$.

Thus, $i_1$ and $i_2$ are not defined at $r=0$. To stress the crucial point again, we want the value of the current $i(r)$ at the point $r=0$ (i.e. $i(0)$), not the limit of the function $i(r)$ as $r$ tends to zero. Hence, the circuit shown in FIG. 1 cannot be solved for the current $i_1$ and $i_2$ by applying the Kirchhoff's laws.

## V. AN ALTERNATE CIRCUIT

In our previous paper [8], we showed another circuit [10] that cannot be solved by applying the Kirchhoff's laws. Elaborating on the method, we shall now show that the circuit shown in FIG. 7 leads to the same kind of indeterminate 0/0 form as the circuit shown in FIG. 1.

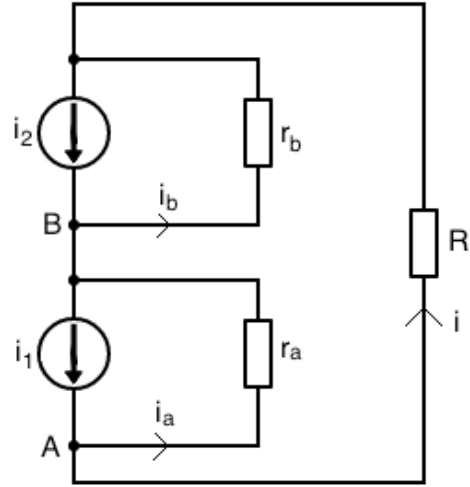

FIG. 7. A circuit to demonstrate the problem slightly differently.

The current sources $i_1$ and $i_2$ are not ideal sources as they have internal resistances $r_a$ and $r_b$ connected in parallel to them, respectively. A constant load resistance $R$ is connected in series with the current sources.

Applying the Kirchhoff's loop law along the outer loop we get,

$$iR = i_a r_a + i_b r_b \qquad (22)$$

Applying the Kirchhoff's current conservation law at nodes "A" and "B", we get

$$i_1 = i + i_a \qquad (23)$$

$$i_2 = i + i_b \qquad (24)$$

Solving these equations for $i_a$ and $i_b$, we get,

$$i_a = \frac{r_b(i_1 - i_2) + i_1 R}{r_a + r_b + R} \qquad (25)$$

$$i_b = \frac{r_a(i_2 - i_1) + i_2 R}{r_a + r_b + R} \qquad (26)$$

At this point, we set $i_1=i_2=i$ and define conductance $c_a = \frac{1}{r_a}$ and $c_b = \frac{1}{r_b}$.

Thus, the values of $i_a$ and $i_b$ reduce to

$$i_a = \frac{iRc_a c_b}{c_a + c_b + Rc_a c_b} \qquad (27)$$

$$i_b = \frac{iRc_a c_b}{c_a + c_b + Rc_a c_b} \quad (28)$$

And we have the voltage across the current source $i_1$ as $v_a = \frac{i_a}{c_a}$ and the voltage across the current source $i_2$ as $v_b = \frac{i_b}{c_b}$.

Substituting the values of $i_a$ and $i_b$ from equations (27) and (28), we get

$$v_a = \frac{iRc_b}{c_a + c_b + Rc_a c_b} \quad (29)$$

$$v_b = \frac{iRc_a}{c_a + c_b + Rc_a c_b} \quad (30)$$

We shall now let the two values of the conductance be equal i.e. $c_a = c_b = c$ (say). We then have,

$$v_a = v_b = \frac{iRc}{2c + Rc^2} \quad (31)$$

Let us denote $v_a$ and $v_b$ as $v(c)$ ('$v$' as a function of '$c$'). We have

$$v(c) = \frac{iRc}{2c + Rc^2} \quad (32)$$

It is evident from the equation (32) that $v(0)$ is undefined. But, the limit of the function $v(c)$ exists, as $c$ tends to zero. This is given by

$$\lim_{c \to 0} \frac{iRc}{2c + Rc^2} = \frac{iR}{2} \quad (33)$$

As seen from the equation (33), the values of the voltages across the ideal current sources of the circuit shown in FIG. 7 attain 0/0 form and cannot be known from the theory. Moreover, the form of the equations is exactly the same except that the currents and the voltages are interchanged (compare equations (21) and (33)).

## VI. CONCLUSIONS AND IMPLICATIONS FOR TEACHING

Armed with the analytical tools presented in this paper, the instructors should be able to explain why the circuit shown in FIG. 1 (and also the circuit referenced in [10]) cannot be completely solved using the existing circuit analysis methods. The new method presented in this paper can also be taught as an alternative method for solving simpler electrical circuits from the solutions of a more general circuit. This should enable the students to see an application of the abstract mathematical concepts like 'limit of a function' in the context of electric circuits. Consider, for example, the circuit shown in FIG. 8.

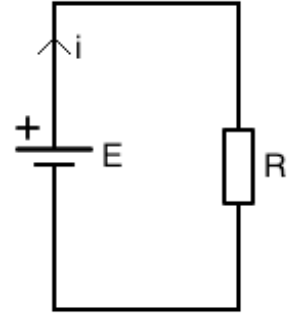

FIG. 8. A general circuit to demonstrate why an ideal voltage source should not be short-circuited.

The current from the voltage source is given by

$$i = \frac{E}{R} \quad (34)$$

As the load resistance $R$ tends to zero, we see that the current from the ideal voltage source $E$ approaches a very large value. This is given by

$$i = \lim_{R \to 0} \frac{E}{R} = \infty \quad (35)$$

As the power drawn from the voltage source approaches a very large value, a constant voltage source should not be connected to a wire of very small resistance.

Consider, as a second example, the circuit shown in FIG. 9.

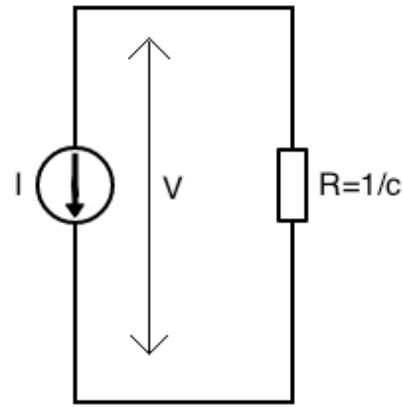

FIG. 9. A general circuit to demonstrate why an ideal current source should not be open-circuited.

The voltage '$V$' across the constant current source '$I$' is given by

$$V = IR \qquad (36)$$

Let us now label the conductance of the resistor as '$c$'. We can then write the equation (36) as

$$V = \frac{I}{c} \qquad (37)$$

As the conductance of the load resistance tends to zero, we can see that the voltage across the constant current source approaches a very large value. This is shown by

$$V = \lim_{c \to 0} \frac{I}{c} = \infty \qquad (38)$$

Hence, the power from the current source approaches a very large value if it is connected across a resistor of a very large value.

Thus, we arrive at the familiar results:

- An ideal voltage source should never be short-circuited and
- An ideal current source should never be open-circuited.

In this way, we hope that the new method introduced in this paper can help the instructors in teaching some fundamental concepts related to the behavior of the voltage and the current sources. However, the primary purpose of introducing the method was to show the reason for breakdown of the Kirchhoff's laws when applied to circuit shown in FIG. 1.

Next, it is interesting to ask if it is possible to setup an experiment to determine the values of $i_1$ and $i_2$ in the circuit shown in FIG. 1. The answer to this question is negative because ideal voltage sources don't exist in the real world. The ideal voltage sources and ideal current sources are mathematical abstractions that will make the analysis of real circuit problems easier. Unfortunately, this important point is not stressed when teaching electric circuit analysis. With the teaching of the breakdown of these abstract ideas as presented in this paper, we believe that the students can better understand their limitations.

## VII. ACKNOWLEDGEMENTS

I should like to thank *Prof. Keith Atkin* for his valuable suggestions and the penetrating comments on the problem. I would also like to thank the three anonymous referees for their invaluable suggestions on the manuscript.